\documentclass[10pt,journal,letterpaper]{IEEEtran}

\usepackage{cite}
\usepackage{amsmath,amssymb,amsfonts}
\usepackage{graphicx}
\usepackage{booktabs}
\usepackage{textcomp}
\usepackage{mathrsfs}

\usepackage{xcolor}
\usepackage{tikz}
\usetikzlibrary{arrows.meta,calc,fit,backgrounds}
\usepackage{pgfplots}
\pgfplotsset{compat=1.18}
\usepgfplotslibrary{groupplots}

\makeatletter
\def\section{\@startsection{section}{1}{\z@}{1.5ex}{0.7ex}{\normalfont\normalsize\centering\scshape}}
\def\subsection{\@startsection{subsection}{2}{\z@}{1.5ex}{0.7ex}{\normalfont\normalsize\itshape}}
\makeatother

\begin{document}

\title{AoGS: A Freshness Metric for Wireless View Scheduling in Dynamic 3D Gaussian Splatting}

\author{Xu~Yang, Branka~Vucetic,~\IEEEmembership{Life Fellow,~IEEE}, and Yonghui~Li,~\IEEEmembership{Fellow,~IEEE}%

\thanks{Xu Yang, Branka Vucetic, and Yonghui Li are with the School of Electrical and Computer Engineering, the University of Sydney, Sydney, Australia. Corresponding author: Xu Yang (xu.yang@sydney.edu.au).}}

\maketitle

\begin{abstract}
Maintaining rendering fidelity in dynamic 3D Gaussian splatting (3DGS) requires timely supervision from multiple viewing directions. Limited wireless uplink resources necessitate efficient view scheduling. To address this challenge, we introduce Age of Gaussian Supervision (AoGS), a metric that tracks, for each Gaussian–direction pair, the elapsed time since the capture of the latest successfully delivered image used for supervision. Before image capture, the server selects cameras to maximize the expected reduction in opacity-weighted AoGS subject to a time-division multiple access (TDMA) airtime budget, accounting for overlapping coverage and probabilistic delivery. To reduce the objective size, we group pairs with identical supervising camera sets, yielding a compact mixed-integer linear programming (MILP) formulation. Under the same airtime budget, both the greedy and MILP-based AoGS schedulers consistently outperform all four baselines in held-out peak signal-to-noise ratio (PSNR) and structural similarity (SSIM) across six tested dynamic scenes. Relative to directional coverage, the MILP-based scheduler reduces mean observable AoGS by 68.5\% and mean learned perceptual image patch similarity (LPIPS) by approximately 10\%, while delivering 16.4\% fewer views per time slot on average.
\end{abstract}

\vspace{-4mm}
\section{Introduction}
\label{intro}
\vspace{-1mm}

Dynamic 3D Gaussian splatting (3DGS) represents evolving scenes with Gaussian ellipsoids, enabling rapid novel-view rendering~\cite{kerbl2023gaussian} for applications such as virtual reality~\cite{fan2025fovgs} and free-view video streaming~\cite{ke2025edgegaussian}. Maintaining rendering quality as the scene evolves requires timely multi-view supervision. In wireless multi-camera systems, however, limited uplink resources can prevent all cameras from delivering their images for every scene update, creating a need for dynamic multi-view scheduling. Active view selection (AVS) methods such as FisherRF~\cite{jiang2024fisherrf} and its extension to dynamic 3DGS~\cite{li2026next} evaluate candidate views using Fisher information, requiring repeated rendering and gradient computation. In wireless scene reconstruction, low-altitude Gaussian splatting~\cite{wang2026lags} selects transmissions based on reconstruction losses computed from captured images, while sample-then-transmit edge Gaussian splatting~\cite{li2026sttgs} spends uplink airtime on pilot images to estimate transmission utilities. These approaches therefore incur additional computational and communication costs. This motivates a scheduling metric that the server can evaluate before image capture without rendering candidate views.

Treating each Gaussian as a virtual monitored source makes Age of Information (AoI)~\cite{yates2021age} a natural foundation for view scheduling. Sensor scheduling based on AoI has been studied in~\cite{kalor2023timely}, which minimizes source AoI when one sensor can update multiple sources. Similarly, one delivered image from a view can supervise multiple Gaussians~\cite{kerbl2023gaussian}. However, repeated supervision from similar directions can keep a Gaussian’s AoI low while supervision from complementary directions becomes stale as the scene evolves. The Gaussian--direction pairs covered by a view depend on the relative geometry of cameras and Gaussians and may change as either moves. This association has been investigated in~\cite{chen2026coverage}, which evaluates how candidate views complement each Gaussian’s previously observed directions. Similarly, ExploreGS~\cite{kim2025exploregs} records observed directions for each Gaussian to guide virtual camera sampling. While Gaussian-level AoI does not distinguish supervision freshness across viewing directions, directional coverage alone does not track how recently each direction was supervised.

Motivated by this need, we introduce Age of Gaussian Supervision (AoGS), which tracks, for each Gaussian--direction pair, the time elapsed since the capture of the most recent successfully delivered image supervising that pair. We derive an exact expression for the expected opacity-weighted AoGS reduction, accounting for overlapping view coverage and probabilistic delivery. The server selects a subset of cameras to maximize this gain under the uplink budget using information available before image capture. We use pattern aggregation to group pairs with identical sets of supervising views, yielding a compact mixed-integer linear programming (MILP) formulation. We evaluate greedy and MILP-based AoGS schedulers on six dynamic scenes over a time-division multiple access (TDMA) uplink and compare them with four baselines based on Gaussian coverage, Gaussian AoI, Gaussian velocity, and directional coverage. Under the same airtime budget, both AoGS schedulers achieve higher held-out peak signal-to-noise ratio (PSNR) and structural similarity (SSIM) than all baselines in every scene, while delivering fewer views per slot on average. Mean learned perceptual image patch similarity (LPIPS) scores range from 0.163 to 0.168 for the baselines, compared with 0.147 for greedy AoGS and 0.146 for MILP-based AoGS. Pattern aggregation reduces the number of objective terms by a factor of 4,146 on average.

\vspace{-1mm}
\section{System Model}
\label{sys}
\vspace{-2mm}

\subsection{Overview}
\vspace{-1mm}

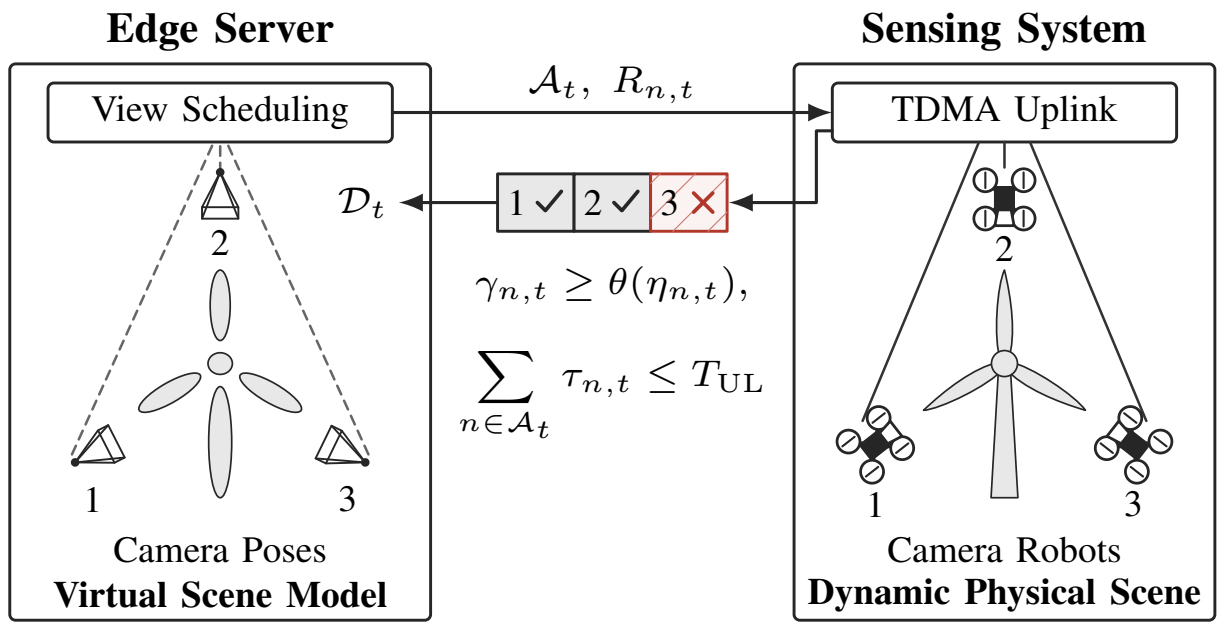
\begin{figure}[t]
    \centering
    \begin{tikzpicture}[
  x=.30pt, y=-.30pt, font=\scriptsize, >=latex,
  line cap=round, line join=round,
  every node/.style={inner sep=0pt, outer sep=0pt},
  frame/.style={draw=black!85, line width=.55pt, rounded corners=1.1pt},
  block/.style={frame, fill=white},
  title/.style={font=\footnotesize\bfseries},
  ray/.style={draw=black!60, line width=.48pt,
              dash pattern=on 1.8pt off 1.3pt},
  uplink/.style={draw=black!80, line width=.48pt},
  flow/.style={->, draw=black!85, line width=.58pt},
  surface/.style={draw=black!85, fill=black!9, line width=.32pt},
  rotor/.style={draw=black!85, fill=white, line width=.52pt},
]
  \definecolor{aogsRed}{HTML}{B2372B}
  \tikzset{
    pics/scene model/.style={code={

      \path[surface] (0,58) ellipse (8.5 and 41);
      \foreach \blade in {0,120,240}
        \path[surface,rotate=\blade] (0,-43) ellipse (7.4 and 26);
      \path[surface] (0,0) ellipse (9 and 8);
    }},
    pics/real scene/.style={code={
      \path[surface] (-9.9,99) -- (9.9,99) -- (4.35,3) -- (-4.35,3) -- cycle;
      \foreach \angle in {0,120,240}
        \path[surface,rotate=\angle] (3,-3)
          .. controls (8.7,-21) and (5.1,-48) .. (0,-69)
          .. controls (-5.4,-58.2) and (-8.4,-18) .. (-3,-3) -- cycle;
      \path[surface] (0,0) circle (9.6);
    }},
    pics/virtual camera/.style={code={
      \draw[black!85,line width=.36pt] (-11.4,21.9) rectangle (11.4,29.1);
      \foreach \corner in {(-11.4,21.9),(11.4,21.9),(-11.4,29.1),(11.4,29.1)}
        \draw[black!85,line width=.36pt] (0,0) -- \corner;
      \fill[black!85] (0,0) circle (2.7);
    }},
    pics/camera robot/.style={code={
      \draw[black!85,line width=.65pt] (-12,-12) -- (12,12) (-12,12) -- (12,-12);
      \foreach \rx/\ry in {-12/-12,12/-12,-12/12,12/12}{
        \begin{scope}[shift={(\rx,\ry)}]
          \path[rotor] (0,0) circle (7.2);
          \draw[black!85,line width=.40pt] (0,-4.9) -- (0,4.9);
        \end{scope}
      }
      \fill[black!85,rounded corners=.45pt] (-6.6,-8.4) rectangle (6.6,7.8);
      \path[rotor] (-4.2,6.6) -- (-6,16.5) -- (6,16.5) -- (4.2,6.6) -- cycle;
    }},
  }

  \foreach \leftedge/\cx/\heading/\process/\cameras/\scene in {
      0/132/Edge Server/View Scheduling/Camera Poses/Virtual Scene Model,
      492/624/Sensing System/TDMA Uplink/Camera Robots/Dynamic Physical Scene}{
    \draw[frame] (\leftedge,32) rectangle ++(264,349);
    \node[title] at (\cx,12) {\heading};
    \draw[block] (\cx-108,44) rectangle ++(216,37);
    \node at (\cx,62.5) {\process};
    \node[font=\fontsize{7}{8.4}\selectfont] at (\cx,337) {\cameras};
    \node[font=\scriptsize\bfseries] at (\cx,365) {\scene};
  }

  \draw[ray] (39,277) -- (128,81) (132,100) -- (132,81) (225,277) -- (136,81);
  \draw[uplink] (535,256) -- (608,81) (624,97) -- (624,81) (713,256) -- (640,81);
  \pic[scale=.85] at (132,220) {scene model};
  \pic[scale=.85] at (624,220) {real scene};
  \foreach \px/\py/\angle in {41/282/117,132/100/0,223/282/-116}
    \pic[rotate=\angle] at (\px,\py) {virtual camera};
  \foreach \px/\py/\angle in {543/271/128,624/117/0,705/271/-128}
    \pic[rotate=\angle] at (\px,\py) {camera robot};
  \foreach \index/\px/\py in {
      1/52/305,2/132/144,3/212/305,1/543/308,2/624/148,3/705/308}
    \node at (\px,\py) {\index};

  \draw[flow] (240,62.5) -- node[above=1.8pt] {$\mathcal{A}_t,\; R_{n,t}$} (516,62.5);
  \draw[flow] (516,74) -- (506,74) -- (506,119) -- (451,119);
  \draw[flow] (305,119) -- (244,119) node[left=2pt] {$\mathcal{D}_t$};
  \foreach \slot/\leftedge in {1/306,2/354,3/402}{
    \begin{scope}[shift={(\leftedge,101)}]
      \ifnum\slot=3
        \path[draw=aogsRed,fill=aogsRed!6,line width=.55pt] (0,0) rectangle (48,36);
        \draw[aogsRed!65,line width=.22pt]
          (0,6)--(6,0) (0,28)--(28,0) (14,36)--(48,2) (36,36)--(48,24);
        \draw[aogsRed,line width=.70pt] (27,12)--(39,24) (27,24)--(39,12);
      \else
        \path[draw=black!85,fill=black!9,line width=.55pt] (0,0) rectangle (48,36);
        \draw[black!85,line width=.70pt] (25,18) -- (30,23) -- (40,11);
      \fi
      \node at (12,18) {\slot};
    \end{scope}
  }

  \node at (378,171) {$\gamma_{n,t} \geq \theta(\eta_{n,t}),$};
  \node at (378,240) {$\displaystyle\sum_{n\in\mathcal{A}_t} \tau_{n,t}
                       \leq T_{\mathrm{UL}}$};
\end{tikzpicture}%
    \vspace{-2mm}
    \caption{Illustration of the system model with three camera views. Each view is represented by its camera pose (left block), which is matched to the corresponding real-world camera carried by a robot (right block). The server schedules all three cameras under the TDMA airtime budget and assigns their transmission rates. Only cameras~1 and~2 deliver their images.}
    \label{fig:system}
    \vspace{-3mm}
\end{figure}

We consider $N$ wireless robots, each equipped with a calibrated camera, observing a dynamic physical scene as illustrated in Fig.~\ref{fig:system}. Time is divided into slots. At each slot $t$, the cameras, indexed by $\mathcal C=\{1,\ldots,N\}$, transmit images to an edge server to supervise updates of a scene representation model $\mathcal S_t$ with $K$ dynamic 3DGS ellipsoids $\mathcal G_{k,t}$ with fixed identities and number as in~\cite{luiten2024dynamic}. The server schedules camera capture and transmission to maintain rendering fidelity under wireless resource constraints, with quality evaluated at held-out viewpoints.

\vspace{-1.5mm}
\subsection{Network Model for View Scheduling}
\label{network}
\vspace{-1mm}

Transmission is over a TDMA uplink. At the start of slot $t$, the server adapts camera transmission rates from channel estimates and schedules views using the current scene state, camera poses, and uplink budget. Scheduled cameras synchronously capture $L$-bit images and transmit sequentially using the full bandwidth $B$. Camera poses are available before scheduling, while metadata overhead and operation delays are neglected.

Assume constant channels within each slot. Let $\gamma_{n,t}$ and $\hat{\gamma}_{n,t}$ denote the actual SNR and the server's estimate, respectively, both in dB. The conditional SNR model is $\gamma_{n,t}\mid\hat{\gamma}_{n,t}\sim\mathcal N(\hat{\gamma}_{n,t},\epsilon^2)$, where $\epsilon$ is the standard deviation of the residual estimation error. Following Gudmundson's shadow-fading correlation model~\cite{gudmundson1991correlation}, each camera's SNR estimates evolve across slots according to a stationary first-order Gauss--Markov process with mean $\bar{\gamma}$, standard deviation $\zeta$, and lag-one correlation $\rho$. Under an ideal outage model, a complete image upload at spectral efficiency $\eta$ succeeds when $\gamma_{n,t}\geq10\log_{10}(2^\eta-1)$, giving the conditional delivery probability
\begin{equation}
p_{n,t}(\eta)
=\Pr\!\left(\gamma_{n,t}\geq\theta(\eta)\mid\hat\gamma_{n,t}\right)
=\Phi\!\left(
\frac{\hat{\gamma}_{n,t}-\theta(\eta)}{\epsilon}
\right),
\label{eq:delivery_probability}
\end{equation}
where $\theta(\eta)=10\log_{10}(2^\eta-1)$ and $\Phi$ is the standard normal cumulative distribution function. Given a finite set $\mathcal M$ of supported spectral efficiencies and reliability target $p_{\min}$, the server selects $\eta_{n,t}=\max\{\eta\in\mathcal M: p_{n,t}(\eta)\geq p_{\min}\}$ for each slot $t$. Cameras without a qualifying rate are ineligible for transmission.

With rate $R_{n,t}=B\eta_{n,t}$ and airtime $\tau_{n,t}=L/R_{n,t}$, the server schedules eligible cameras $\mathcal A_t\subseteq\mathcal C$ satisfying $\sum_{n\in\mathcal A_t}\tau_{n,t}\leq T_{\mathrm{UL}}$, where $T_{\mathrm{UL}}$ is the per-slot uplink budget. Each upload is attempted once and consumes its allocated airtime regardless of outcome. Conditional on the channel estimates, uploads succeed independently across cameras with probabilities $p_{n,t}=p_{n,t}(\eta_{n,t})$. At slot end, received images indexed by $\mathcal D_t\subseteq\mathcal A_t$ supervise the scene update.

\vspace{-1.5mm}
\subsection{Scene Representation and Updating}
\vspace{-1mm}

The server maintains a scene representation model $\mathcal S_t=\{\mathcal G_{k,t}\}_{k=1}^{K}$, where $\mathcal G_{k,t}=(\boldsymbol\mu_{k,t},\mathbf q_{k,t}, \mathbf s_k,o_k,\mathbf c_k)$ comprises the center $\boldsymbol\mu_{k,t}$, rotation quaternion $\mathbf q_{k,t}$, scales $\mathbf s_k$, opacity logit $\mathbf o_k$, and RGB color $\mathbf c_k$. Gaussian identities, $K$, $\mathbf s_k$, $o_k$, and $\mathbf c_k$ remain fixed after initialization, with opacity $\sigma(o_k)=(1+e^{-o_k})^{-1}$. Rendering uses differentiable Gaussian projection and compositing. $\boldsymbol\mu_{k,t}$ and $\mathbf q_{k,t}$ are parameters to be optimized during each slot $t$.

Gaussian parameters and camera poses determine supervision coverage before scheduling. Delivered images $\mathcal I_t^{\mathrm{rx}}=\{I_{n,t}:n\in\mathcal D_t\}$ then update the model:
\begin{equation}
\mathcal S_t^{+}
=\Psi(\mathcal S_t,\mathcal I_t^{\mathrm{rx}}),
\end{equation}
where $\Psi$ optimizes centers and rotations using reconstruction losses and temporal regularization from~\cite{luiten2024dynamic}. If $\mathcal D_t=\varnothing$, optimization is skipped and $\mathcal S_t^{+}=\mathcal S_t$. The resulting model is used in the next slot. Thus, the reconstruction benefit of a candidate schedule depends on delivered images unavailable at scheduling time, motivating a surrogate that can be evaluated before capture. 

\section{Age of Gaussian Supervision}
\vspace{-1mm}

The proposed Age of Gaussian Supervision (AoGS) tracks supervision freshness for each Gaussian--direction pair. The server computes the expected aggregate AoGS reduction from view coverage, supervision history, and channel estimates before image capture. We use this quantity as a scheduling surrogate under the TDMA airtime constraint.

\vspace{-1.5mm}
\subsection{Directional Supervision Coverage}
\vspace{-1mm}

\begin{figure}[t]
    \centering
    \input{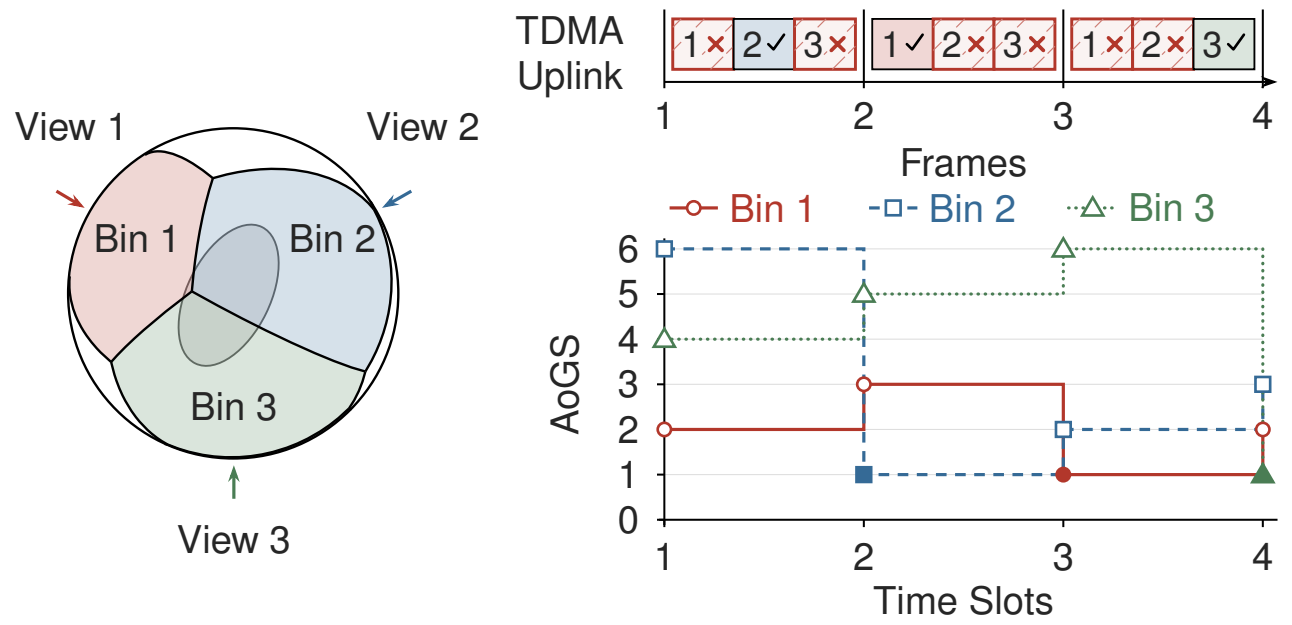}
    \vspace{-2mm}
    \caption{Illustration of AoGS for one Gaussian using eight Fibonacci bins. (a) Camera views provide supervision through distinct direction bins. (b) AoGS evolution based on TDMA delivery, with all three views scheduled in each slot. Different initial ages are used for clarity.}
    \label{fig:aogs}
    \vspace{-3mm}
\end{figure}

At the beginning of slot $t$, the server computes directional coverage from the current scene model and camera poses before scheduling. To distinguish supervision from different viewing directions, we partition each Gaussian's viewing directions into $|\mathcal B|$ non-overlapping bins $\{\Omega_b\}_{b\in\mathcal B}$. The bins are defined using fixed coordinate axes shared by the entire scene, so their orientations do not change when the Gaussian rotates. We generate approximately uniform bin centers using spherical Fibonacci sampling~\cite{keinert2015spherical} and assign each direction to its nearest center, as illustrated in Fig.~\ref{fig:aogs} (a). Let $\boldsymbol\nu_{n,t}$ denote the center of camera $n$. The unit direction from camera $n$ to Gaussian $k$ is $\mathbf d_{nk,t}= \frac{\boldsymbol\mu_{k,t}-\boldsymbol\nu_{n,t}} {\|\boldsymbol\mu_{k,t}-\boldsymbol\nu_{n,t}\|_2}$.

Let $v_{nk,t}\in\{0,1\}$ indicate the visibility of Gaussian $k$ from camera $n$. Under our projection-based approximation, $v_{nk,t}=1$ if and only if the Gaussian center lies within the camera's viewing frustum and the Gaussian's opacity satisfies $\sigma(o_k)\geq\alpha_{\min}$. The directional coverage indicator for camera $n$ and Gaussian--direction pair $(k,b)$ is
\begin{equation}
    h_{nkb,t}
    =
    v_{nk,t}\,
    \mathbf 1\{\mathbf d_{nk,t}\in\Omega_b\}.
    \label{eq:directional_coverage}
\end{equation}
Thus, each camera can cover many Gaussian--direction pairs but at most one direction bin per Gaussian. Different cameras may cover the same pair or observe the same Gaussian from different bins.

The pairs covered by each camera can change across slots as the Gaussians or cameras move. After transmission, the server uses the images received from cameras in $\mathcal D_t$ to update the scene model. Under this coverage model, a Gaussian--direction pair $(k,b)$ receives supervision in slot $t$ if and only if $h_{nkb,t}=1$ for at least one camera $n\in\mathcal D_t$.

\vspace{-1.5mm}
\subsection{AoGS Evolution}
\vspace{-1mm}

For each Gaussian--direction pair, AoGS measures the time elapsed since capture of the most recent image used to supervise that pair. For a pair $(k,b)$ with prior supervision, let $u_{kb}(t)$ denote the capture slot of its most recent successfully delivered supervising image as of the beginning of slot $t$. Its AoGS at that time is
\begin{equation}
    \Delta_{kb}(t)=t-u_{kb}(t).
    \label{eq:aogs_definition}
\end{equation}
Scheduling begins at $t=1$, with $\Delta_{kb}(1)=1$ for pairs supervised during initialization. For pairs with no prior supervision, we assign an initial age of $\Delta_{kb}(1)=2$.

After transmission in slot $t$, the server uses the directional coverage indicators in~\eqref{eq:directional_coverage} and the set of cameras with successful deliveries, $\mathcal D_t$, to determine whether pair $(k,b)$ received supervision:
\begin{equation}
    U_{kb,t}
    =
    \mathbf 1\!\left\{
        \sum_{n\in\mathcal D_t}h_{nkb,t}>0
    \right\}.
    \label{eq:aogs_supervision}
\end{equation}
Thus, $U_{kb,t}=1$ if and only if at least one camera in $\mathcal D_t$ covers the pair. The AoGS at the beginning of slot $t+1$ is
\begin{equation}
    \Delta_{kb}(t+1)
    =
    1+\bigl(1-U_{kb,t}\bigr)\Delta_{kb}(t).
    \label{eq:aogs_evolution}
\end{equation}
A pair supervised in slot $t$ starts slot $t+1$ with age one. Otherwise, its age increases by one. Multiple delivered images covering the same pair produce a single age update. AoGS therefore preserves the directional distinction between recent and stale supervision while accounting for overlapping camera coverage. An example of AoGS evolution is in Fig.~\ref{fig:aogs} (b).

\section{Problem Formulation}
\vspace{-1mm}

\subsection{AoGS-Based View Scheduling}
\vspace{-1mm}

At the beginning of slot $t$, the server evaluates candidate camera subsets using the current AoGS values $\Delta_{kb}(t)$, directional coverage indicators $h_{nkb,t}$, and channel estimates $\hat{\gamma}_{n,t}$. Consider a candidate schedule $\mathcal A\subseteq\mathcal C$ containing only cameras with a qualifying rate under the rule in Section~\ref{network}. Given the channel estimates, uploads succeed independently with probabilities $p_{n,t}$ obtained from~\eqref{eq:delivery_probability} at the selected spectral efficiencies $\eta_{n,t}$. Hence, the probability that Gaussian--direction pair $(k,b)$ receives supervision under $\mathcal A$ is
\begin{equation}
    q_{kb,t}(\mathcal A)
    =1-\prod_{n\in\mathcal A}
    \bigl(1-p_{n,t}h_{nkb,t}\bigr),
    \label{eq:aogs_refresh_probability}
\end{equation}
where the product term accounts for the case where no scheduled camera successfully delivers an image covering the pair. Scheduling an additional camera that covers the pair can increase the supervision probability, while multiple successful deliveries still produce a single freshness update.

By~\eqref{eq:aogs_evolution}, the next-slot age is $\Delta_{kb}(t)+1$ when $U_{kb,t}=0$ and one when $U_{kb,t}=1$. We weight Gaussian $k$ by its opacity $\sigma(o_k)$ and define the expected aggregate AoGS reduction relative to receiving no supervision as
\begin{equation}
\begin{aligned}
G_t(\mathcal A)
&= \sum_{k,b} \sigma(o_k) \Bigl[\Delta_{kb}(t)+1-
\mathbb E\!\left[
\Delta_{kb}(t+1)
\,\middle|\,
\substack{
\mathcal A,\;\Delta_{kb}(t),\\
\{h_{nkb,t}\}_{n},\\ \{\hat\gamma_{n,t}\}_{n}
}
\right]\Bigr] \\
&= \sum_{k,b} \sigma(o_k) \Delta_{kb}(t)
\mathbb E\!\left[
U_{kb,t}
\,\middle|\,
\substack{
\mathcal A,\;\Delta_{kb}(t),\\
\{h_{nkb,t}\}_{n},\\ \{\hat\gamma_{n,t}\}_{n}
}
\right] \\
&= \sum_{k,b} \sigma(o_k) \Delta_{kb}(t)
q_{kb,t}(\mathcal A) \\
&= \sum_{k,b} \sigma(o_k) \Delta_{kb}(t)
\Bigl[
1-\prod_{n\in\mathcal A}
\bigl(1-p_{n,t}h_{nkb,t}\bigr)
\Bigr].
\end{aligned}
\label{eq:aogs_expected_gain}
\end{equation}
Since $U_{kb,t}$ is a binary indicator, its conditional expectation equals $q_{kb,t}(\mathcal A)$. Linearity of expectation allows the contributions to be summed even when different pairs receive supervision from the same image.

The server selects the schedule $\mathcal A_t$ by maximizing this expected reduction under the TDMA airtime budget
\begin{equation}
\begin{aligned}
    \underset{\mathcal A\subseteq\mathcal C}{\operatorname{maximize}}
    \quad & G_t(\mathcal A)\\
    \operatorname{subject\ to}
    \quad & \sum_{n\in\mathcal A}\tau_{n,t}
    \leq T_{\mathrm{UL}}.
\end{aligned}
\label{eq:aogs_scheduling}
\end{equation}
Since the current ages and opacity weights are fixed when scheduling, maximizing $G_t(\mathcal A)$ is equivalent to minimizing the expected opacity-weighted aggregate AoGS at the beginning of slot $t+1$. The objective balances supervision freshness, directional coverage, and delivery reliability, while the TDMA constraint determines which combinations of camera uploads are feasible.

\vspace{-1.5mm}
\subsection{Supervision-Pattern Aggregation}
\vspace{-1mm}

The objective in~\eqref{eq:aogs_expected_gain} can easily involve millions of Gaussian--direction contributions in real-world scenes, resulting in a large optimization formulation. However, for any candidate schedule $\mathcal A$, pairs covered by the same set of cameras have the same probability of receiving supervision. Motivated by the aggregation of demand points with identical facility coverage in maximal covering formulations~\cite{lv2025presolving}, we propose a pattern aggregation algorithm to group Gaussian--direction pairs by their covering cameras to obtain a compact objective.

For each pair $(k,b)$, define its supervision pattern as the set of cameras that cover it at slot $t$,
\begin{equation}
    \mathcal V_{kb,t}
    = \{n\in\mathcal C:h_{nkb,t}=1\}.
    \label{eq:supervision_pattern}
\end{equation}
Let $\mathscr V_t$ denote the set of distinct nonempty patterns. For each $\mathcal V\in\mathscr V_t$, the pairs in $\mathcal E_{\mathcal V,t} =\{(k,b):\mathcal V_{kb,t}=\mathcal V\}$ have total weight
\begin{equation}
    W_{\mathcal V,t}
    =
    \sum_{(k,b)\in\mathcal E_{\mathcal V,t}}
    \sigma(o_k)\Delta_{kb}(t).
    \label{eq:pattern_weight}
\end{equation}
Factoring out their common supervision probability gives the gain for any candidate schedule $\mathcal A$,
\begin{equation}
    G_t(\mathcal A)
    =
    \sum_{\mathcal V\in\mathscr V_t}
    W_{\mathcal V,t}
    \left[
    1-\prod_{n\in\mathcal A\cap\mathcal V}(1-p_{n,t})
    \right].
    \label{eq:aggregated_aogs_gain}
\end{equation}
Pairs with empty patterns contribute zero. Thus, \eqref{eq:aggregated_aogs_gain} equals \eqref{eq:aogs_expected_gain} for every feasible schedule and therefore preserves the optimal value and all optimal schedules.

To linearize the products in~\eqref{eq:aggregated_aogs_gain}, let $x_n\in\{0,1\}$ indicate whether camera $n$ is scheduled. For each pattern $\mathcal V$, let $n_i$ denote the $i$th camera in an arbitrary ordering of its cameras. Let $r_{\mathcal V,i}$ denote the probability that none of the first $i$ cameras successfully delivers an image. Under the TDMA airtime budget, the equivalent MILP is
\begin{equation}
\label{eq:aogs_milp}
\begin{aligned}
\max_{\boldsymbol{x},\boldsymbol{r}}\quad
& \sum_{\mathcal V\in\mathscr V_t}
  W_{\mathcal V,t}(1-r_{\mathcal V,|\mathcal V|}) \\
\mathrm{s.t.}\quad
& \sum_{n\in\mathcal C}\tau_{n,t}x_n\le T_{\mathrm{UL}},\\
& (1-p_{n_i,t})r_{\mathcal V,i-1}
  \le r_{\mathcal V,i}\le r_{\mathcal V,i-1},\\
& r_{\mathcal V,i}\ge
  r_{\mathcal V,i-1}-p_{n_i,t}x_{n_i},\\
& r_{\mathcal V,i}\le
  (1-p_{n_i,t})r_{\mathcal V,i-1}
  +p_{n_i,t}(1-x_{n_i}),\\
& 0\le r_{\mathcal V,i}\le1,\\
& r_{\mathcal V,0}=1, \qquad x_n\in\{0,1\}.
\end{aligned}
\end{equation}
Here, $n\in\mathcal C$, $\mathcal V\in\mathscr V_t$, and $i=1,\ldots,|\mathcal V|$. Since $x_{n_i}$ is binary, the constraints enforce $r_{\mathcal V,i}=r_{\mathcal V,i-1}(1-p_{n_i,t}x_{n_i})$. Hence, $1-r_{\mathcal V,|\mathcal V|}$ equals the supervision probability for pattern $\mathcal V$ in~\eqref{eq:aggregated_aogs_gain}.

The aggregated objective has one term per distinct nonempty pattern, with $|\mathscr V_t|\le\min\{K|\mathcal B|,2^N-1\}$. The server updates pattern membership when changes in scene geometry or camera poses alter coverage. At each slot, it computes the pattern weights from the current pair ages, which evolve individually according to~\eqref{eq:aogs_evolution}.

\section{Numerical Results}
\vspace{-1mm}

\subsection{Simulation Setup}
\vspace{-1mm}

We conduct simulations on six scenes from the PanopticSports dataset~\cite{luiten2024dynamic}. Each scene contains 150 frames, with 27 candidate and 4 held-out views. Each frame corresponds to a simulated time slot. The first slot $t=0$ initializes the model using all candidate views in $\mathcal C$ and runs $J_{\mathrm{init}}=10{,}000$ optimization iterations without imposing TDMA budget. In each of the remaining 149 slots $t$, we schedule camera uploads and update the model for $J_{\mathrm{upd}}=500$ iterations using only the delivered images in $\mathcal I_t^{\mathrm{rx}}$. If no image is delivered, we skip the update. We use $|\mathcal B|=32$ direction bins and $\alpha_{\min}=0.01$. For the SNR model, we set $\bar{\gamma}=18$~dB, $\zeta=5$~dB, $\epsilon=2$~dB, and $\rho=0.9$. The uplink parameters are $B=4$~MHz, $L=180$~kbit, $T_{\mathrm{UL}}=50$~ms, and $p_{\min}=0.95$, with supported spectral efficiencies $\mathcal M=\{0.25,0.5,1,1.5,2,3,4,5,6\}$~bit/s/Hz.

Our proposed scheduler, AoGS-E, uses HiGHS to solve the MILP in~\eqref{eq:aogs_milp} for the aggregated expected AoGS gain. We compare it with four greedy baselines. Gaussian coverage (GC) uses opacity-weighted Gaussian coverage without distinguishing viewing directions. Gaussian age (GA) additionally weights each Gaussian by its AoI~\cite{yates2021age}, measured as the time since capture of the latest delivered image used to supervise it. Velocity (VEL) replaces the age weight with the Gaussian's estimated speed. Directional coverage (DC), inspired by~\cite{chen2026coverage}, uses opacity-weighted Gaussian--direction coverage without age weighting. To assess the effect of greedy selection, we also evaluate AoGS-G, which applies the greedy rule to the same expected gain, expressed before aggregation in~\eqref{eq:aogs_expected_gain}.

All five greedy methods use the same selection rule. At each slot, they consider cameras with positive delivery probability whose uploads fit within the remaining airtime budget. They select the camera with the largest expected marginal gain per unit airtime. If all eligible cameras have zero marginal gain, they select the one requiring the least airtime. Selection continues until no eligible camera remains. Within each scene, all methods use the same initialized model, candidate views, channel traces, random draws for delivery outcomes, airtime budget, and number of optimization iterations per update.

After each scene update, we evaluate reconstruction quality on the four held-out views using peak signal-to-noise ratio (PSNR) for RGB intensities normalized to $[0,1]$, structural similarity index (SSIM)~\cite{wang2004image}, and learned perceptual image patch similarity
(LPIPS)~\cite{zhang2018perceptual} with an AlexNet backbone. Higher PSNR and SSIM values and lower LPIPS values indicate better reconstruction. We also report mean observable AoGS, computed for slot $t$ as the opacity-weighted mean of the next-slot ages $\Delta_{kb}(t+1)$ over Gaussian--direction pairs covered by at least one candidate camera before scheduling. Reconstruction scores are averaged over four held-out views, and all metrics are averaged equally over time slots and scenes.

\vspace{-1.5mm}
\subsection{Performance Results}
\vspace{-1mm}

\begin{table*}[t]
\centering
\caption{Per-scene reconstruction quality: PSNR $\uparrow$ / SSIM $\uparrow$.}
\label{tab:reconstruction_quality}
\vspace{-1mm}

\footnotesize
\setlength{\tabcolsep}{1.5pt}
\renewcommand{\arraystretch}{1.18}

\begin{tabular*}{\textwidth}{@{\extracolsep{\fill}}l*{7}{c}@{}}
\toprule
\textbf{Method}
& \textbf{Basketball}
& \textbf{Boxes}
& \textbf{Football}
& \textbf{Juggle}
& \textbf{Softball}
& \textbf{Tennis}
& \textbf{Mean} \\
\midrule
GC
& 27.22\,/\,0.866
& 27.59\,/\,0.882
& 27.06\,/\,0.875
& 28.50\,/\,0.894
& 27.05\,/\,0.886
& 28.04\,/\,0.890
& 27.58\,/\,0.882 \\
GA
& 27.30\,/\,0.868
& 27.68\,/\,0.883
& 27.24\,/\,0.879
& 28.53\,/\,0.895
& 27.10\,/\,0.887
& 28.09\,/\,0.892
& 27.66\,/\,0.884 \\
VEL
& 27.39\,/\,0.870
& 27.62\,/\,0.884
& 27.19\,/\,0.880
& 28.54\,/\,0.894
& 27.09\,/\,0.887
& 28.13\,/\,0.892
& 27.66\,/\,0.885 \\
DC
& 27.41\,/\,0.870
& 27.85\,/\,0.887
& 27.30\,/\,0.880
& 28.62\,/\,0.897
& 27.18\,/\,0.889
& 28.18\,/\,0.894
& 27.76\,/\,0.886 \\
\midrule
AoGS-G
& 27.55\,/\,0.877
& 28.02\,/\,\textbf{0.894}
& 27.51\,/\,0.889
& 28.73\,/\,0.904
& 27.32\,/\,0.896
& \textbf{28.32}\,/\,\textbf{0.900}
& 27.91\,/\,0.893 \\
AoGS-E
& \textbf{27.68}\,/\,\textbf{0.879}
& \textbf{28.06}\,/\,\textbf{0.894}
& \textbf{27.59}\,/\,\textbf{0.890}
& \textbf{28.80}\,/\,\textbf{0.905}
& \textbf{27.34}\,/\,\textbf{0.897}
& \textbf{28.32}\,/\,\textbf{0.900}
& \textbf{27.97}\,/\,\textbf{0.894} \\
\bottomrule
\end{tabular*}
\end{table*}

\begin{figure}[t]
    \centering
    \includegraphics[width=\columnwidth]{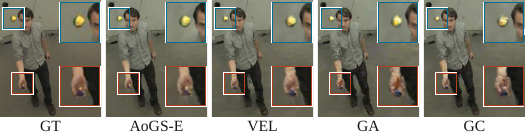}
    \vspace{-6mm}
    \caption{Rendering comparison for the same frame and held-out view in the Juggle scene. GT denotes ground truth.}
    \vspace{-3mm}
    \label{fig:juggle_render_comparison}
\end{figure}

\subsubsection{Per-Scene Reconstruction}

Table~\ref{tab:reconstruction_quality} compares reconstruction quality across six scenes under the same uplink airtime budget. AoGS-E achieves the best or tied-best PSNR and SSIM in every scene. Compared with DC, it improves the mean PSNR from $27.76$ to $27.97$~dB and mean SSIM from $0.886$ to $0.894$. AoGS-G also outperforms all four baselines on both metrics in every scene, supporting the benefit of incorporating directional freshness into view selection even with a greedy scheduler. AoGS-E further increases the mean PSNR by $0.06$~dB over AoGS-G.

Fig.~\ref{fig:juggle_render_comparison} compares the images rendered by AoGS-E, VEL, GA, and GC with the ground-truth image for the Juggle scene. In the enlarged regions, AoGS-E most clearly preserves the outlines of the yellow ball and the hand holding the second ball. The other methods show more color noise and jagged edges that distort the shapes of the ball and hand.

\subsubsection{AoGS-Fidelity Relationship}

\begin{table}[t]
\centering
\caption{Mean LPIPS, Observable AoGS, Gaussian AoI, and numbers of scheduled and delivered views per time slot across six scenes.}
\label{tab:aogs_effect}
\vspace{-2mm}
\footnotesize
\setlength{\tabcolsep}{2pt}
\renewcommand{\arraystretch}{1.15}

\begin{tabular*}{\columnwidth}{@{\extracolsep{\fill}}lccccc@{}}
\toprule
\textbf{Method}
& \textbf{LPIPS} $\downarrow$
& \textbf{AoGS} $\downarrow$
& \textbf{AoI} $\downarrow$
& \textbf{Scheduled}
& \textbf{Delivered} \\
\midrule
GC     & 0.168 & 12.85 & 1.33          & 4.72 & 4.64 \\
GA     & 0.165 & 11.94 & \textbf{1.17} & 4.57 & 4.49 \\
VEL    & 0.163 & 11.01 & 1.46          & 4.83 & 4.76 \\
DC     & 0.163 & 11.60 & 1.83          & 4.93 & 4.83 \\
\midrule
AoGS-G & 0.147          & 3.77          & 1.36 & 4.06 & 3.98 \\
AoGS-E & \textbf{0.146} & \textbf{3.65} & 1.35 & 4.12 & 4.04 \\
\bottomrule
\end{tabular*}
\end{table}

Table~\ref{tab:aogs_effect} shows that AoGS-E and AoGS-G achieve both lower mean observable AoGS and lower held-out LPIPS than all four baselines under the same airtime budget. Compared with DC, AoGS-E reduces mean observable AoGS from $11.60$ to $3.65$ and held-out LPIPS from $0.163$ to $0.146$, while delivering fewer views per time slot ($4.04$ versus $4.83$). AoGS-G achieves similar fidelity, with a mean observable AoGS of $3.77$ and an LPIPS of $0.147$. GA achieves the lowest mean Gaussian AoI ($1.17$), yet its mean observable AoGS is $11.94$. A new image from a recently supervised direction can reset a Gaussian's AoI even as supervision from other directions grows stale.

\subsubsection{Effect of Pattern Aggregation}

\begin{table}[t]
\centering
\caption{Number of objective terms per time slot before and after pattern aggregation.}
\label{tab:pattern_aggregation}
\vspace{-2mm}
\footnotesize
\setlength{\tabcolsep}{2pt}
\renewcommand{\arraystretch}{1.15}

\begin{tabular*}{\columnwidth}{
    @{\extracolsep{\fill}}lccc@{}
}
\toprule
\textbf{Scene}
& \shortstack{\textbf{Covered}\\\textbf{pairs}}
& \shortstack{\textbf{Aggregated}\\\textbf{patterns}}
& \textbf{Reduction} \\
\midrule
Basketball & 2,247,015 & 450.7 & $4{,}986\times$ \\
Boxes      & 1,796,848 & 456.4 & $3{,}937\times$ \\
Football   & 1,835,170 & 448.6 & $4{,}091\times$ \\
Juggle     & 1,908,375 & 454.3 & $4{,}201\times$ \\
Softball   & 1,755,324 & 450.8 & $3{,}894\times$ \\
Tennis     & 1,697,262 & 450.3 & $3{,}769\times$ \\
\bottomrule
\end{tabular*}
\end{table}

Table~\ref{tab:pattern_aggregation}
compares the number of Gaussian--direction pairs with nonempty coverage and the number of objective terms before and after aggregation for the same AoGS-E inputs in each time slot. The counts are averaged over the 149 time slots in each scene. Across the six scenes, approximately $1.87$ million covered pairs are represented by $452$ distinct supervision patterns on average. The mean reduction factor ranges from $3{,}769\times$ to $4{,}986\times$, with an overall mean of $4{,}146\times$. This reduction preserves the objective in~\eqref{eq:aggregated_aogs_gain}
because all pairs in a pattern share the same supervision probability under any schedule, allowing their opacity-weighted ages to be summed.

\vspace{-1mm}
\section{Conclusion}
\vspace{-1mm}

This paper proposed AoGS-based scheduling for wireless dynamic 3DGS reconstruction, using the supervision age of each Gaussian--direction pair to allocate limited uplink airtime. Pattern aggregation preserves the scheduling objective while enabling a compact MILP formulation. Across six scenes, AoGS-E reduced mean observable AoGS by $68.5\%$ and LPIPS by approximately $10\%$ relative to directional coverage (DC), while delivering $4.04$ views per time slot on average compared with $4.83$ for DC. Aggregation reduced the number of objective terms by a factor of $4{,}146$ on average.

\bibliographystyle{IEEEtran}
\bibliography{reference}

\end{document}